\documentclass[10pt]{article}
\usepackage[utf8]{inputenc}
\usepackage[margin=0.8in]{geometry}
\usepackage{epsfig,amssymb,soul,url,cite,graphicx,epstopdf,amsfonts,amsmath,bm,setspace,comment,siunitx,mathrsfs,multirow,mathtools,array,lipsum,enumitem,ifthen,leftidx,relsize,braket,float,color,titlesec,microtype,caption}
\usepackage[normalem]{ulem}

\usepackage[breaklinks=true,colorlinks=true,hypertexnames=false,linktoc=section,pdfencoding=auto,psdextra=true,linkcolor=blue,anchorcolor=black,citecolor=blue,filecolor=black,menucolor=black,runcolor=black,urlcolor=blue]{hyperref}
\usepackage[affil-it]{authblk}
\usepackage[dvipsnames]{xcolor}
\usepackage[switch]{lineno}

\usepackage{mathpazo}
\usepackage[semibold]{sourcesanspro}
\DeclareRobustCommand{\sbseries}{\fontseries{sb}\selectfont}
\DeclareTextFontCommand{\textsb}{\sbseries}
\titleformat*{\section}{\large\sffamily\sbseries}
\titleformat*{\subsection}{\large\sffamily\sbseries}
\makeatletter
\def\title@font{\Large\sbseries}
\let\ltx@maketitle\@maketitle
\def\@maketitle{\bgroup%
	\let\ltx@title\@title%
	\def\@title{\resizebox{\textwidth}{!}{%
			\mbox{\title@font\ltx@title}%
	}}%
	\ltx@maketitle%
	\egroup}
\makeatother

\title{On the role of the microstructure in the deformation of porous solids}
\author{Sansit Patnaik}
\author{Mehdi Jokar}
\author{Wei Ding}
\author{Fabio Semperlotti \thanks{To whom correspondence should be addressed. Email: fsemperl@purdue.edu}}
\affil{Ray W. Herrick Laboratories, School of Mechanical Engineering, Purdue University\\West Lafayette, Indiana, USA - 47907}

\begin{document}
\date{}
\maketitle
\section*{Abstract}
\noindent\textsb{This study explores the role that the microstructure plays in determining the macroscopic static response of porous elastic continua and exposes the occurrence of position-dependent nonlocal effects that are strictly correlated to the configuration of the microstructure. Then, a nonlocal continuum theory based on variable-order fractional calculus is developed in order to accurately capture the complex spatially distributed nonlocal response. The remarkable potential of the fractional approach is illustrated by simulating the nonlinear thermoelastic response of porous beams. The performance, evaluated both in terms of accuracy and computational efficiency, is directly contrasted with high-fidelity finite element models that fully resolve the pores' geometry. Results indicate that the reduced-order representation of the porous microstructure, captured by the synthetic variable-order parameter, offers a robust and accurate representation of the multiscale material architecture that largely outperforms classical approaches based on the concept of average porosity.}

\subsection*{{\normalsize{Porous solids | Nonlocal effects | Variable-order fractional calculus}}}

\section*{Introduction}
Porous materials have a long and distinguished history of applications spanning the most diverse fields of science and engineering \cite{day2021evolution,zou2017porous,krishna2008engineered}. In nature, the porous architecture is pervasive particularly in those applications that require the simultaneous achievement of low weight and high stiffness; a concept known as high specific stiffness. The most notable and broadly available example of such material is found in animal bones, although other compelling examples of naturally occurring porous materials include wood, marine shells, and rocks \cite{fellah2004ultrasonic,mannan2017correlations,mannan2018stiffness}. The pursuit of materials endowed with extreme specific stiffness has long been a major quest in structural engineering. Learning from nature, it would appear logical to make more extensive use of porous architectures, hence providing transformational alternatives to more traditional materials like alloys and composites. While the design of nature-inspired engineering materials is not new in the scientific community and has seen many successful applications \cite{zhao2014bioinspired}, the use of porous materials within the design cycle has not progressed accordingly. The main reasons contributing to this lag are rooted in either technical or theoretical aspects. At a technical level, porous materials are extremely challenging to fabricate and the few realizations typically rely on a random microstructure (e.g. metal foams \cite{gibson2000mechanical}). At a theoretical level, the incomplete understanding of how size effects emerge from the underlying microstructure and affect the intrinsic multiscale response of the material limits significantly the ability to design and simulate their performance. The recent rapid advances in additive manufacturing have markedly eased the technical concerns, hence suggesting that the design of engineered structural materials with carefully crafted porosity could be finally in reach. 

On the contrary, the theoretical gap is still to be vanquished, hence severely hindering the use of porosity as a design concept. Looking more closely at the theoretical aspects, several theoretical and experimental investigations have shown the central role that size effects play in determining the response of porous structures as well as nontrivial phenomena such as softening behavior \cite{chouksey2019computational,laubie2017stress}, and power-law scaling of structural constitutive relations (e.g. dispersion relations, effective structural strength, and average permeability \cite{fellah2000transient,fellah2004ultrasonic,grosman2008influence,hellstrom2010mechanisms,mannan2018stiffness}). However, existing models do not accurately capture the size effects because they adopt homogenized microstructural descriptions that oversimplify the effect of the porous geometry by capturing only the degree of porosity, while ignoring the size and distribution of the pores \cite{patnaik2021porous,laubie2017stress}. From a practical perspective, the homogenization process discards many physical attributes of the porous microstructure when modeling its size effects within the continuum. Indeed, existing models either fully ignore size effects (e.g. rule of mixture models \cite{madsen2003physical,anirudh2019comprehensive}) or assume an \textit{ad hoc} spatially-uniform distribution of the size effects (e.g. classical nonlocal and micro-mechanical models \cite{batra159misuse}). We will show that, at the continuum level, this latter assumption leads to an inaccurate physical representation of the effect of the porous microstructure. From a more general perspective, and independently of the specific approach, existing modeling strategies share a common set of critical limitations including: [L1] the demand for computationally prohibitive resources, when attempting to fully resolve the pores' geometry in pursue of high-fidelity approaches; and [L2] the lack of accuracy in numerical predictions, typically driven by the inability to account for the effect of the pores' configuration and their resulting impact on size effects. The ability to overcome these critical limitations would unlock the many opportunities offered by the broad class of porous engineered materials, hence enabling a fundamental leap forward in a variety of pivotal applications including, but not limited to, light-weight components for aerospace, terrestrial, and marine transportation \cite{goransson2008tailored,xue2019structural}, batteries and energy storage systems \cite{zhao2016understanding,li2022dynamics}, and even robotics and biomedical implants \cite{krishna2008engineered}.

In this study, we attempt to bridge the fundamental theoretical gaps highlighted by the above limitations L1 and L2. We start by embracing the intrinsically nonlocal nature of porous solids by directly linking the size effects to the specific characteristics of the porous microstructure. This approach provides a solid physical foundation to develop highly accurate and computationally efficient nonlocal continuum models and it opens the way to new generation theoretical models based on variable-order (VO) fractional calculus\footnote{VO operators are differ-integral operators that allow the order of differentiation (or integration) to vary as a function of either dependent or independent variables \cite{samko1993integration}. This unique feature opens immense possibilities for the modeling of several complex phenomena including, for example, anomalous transport \cite{chechkin2005fractional,sokolov2006field} and viscoelasticity \cite{di2020novel,suzuki2021fractional}; for a detailed review, please refer to \cite{patnaik2020review}.}. Notably, the use of VO operators is not a mere mathematical expedient to more conveniently model these complex media, but instead it is a natural consequence of the nature of the underlying microstructure. In other terms, we will show that the fractional-order mechanics emerges naturally from the physical characteristics of porous media so that the resulting model can accurately resolve the interplay between the porous geometry, the size effects, and the experimentally observed power-law characteristics \cite{west2016fractional,patnaik2020generalized}.

\begin{figure}[ht]
	\centering
	\includegraphics[width=\linewidth]{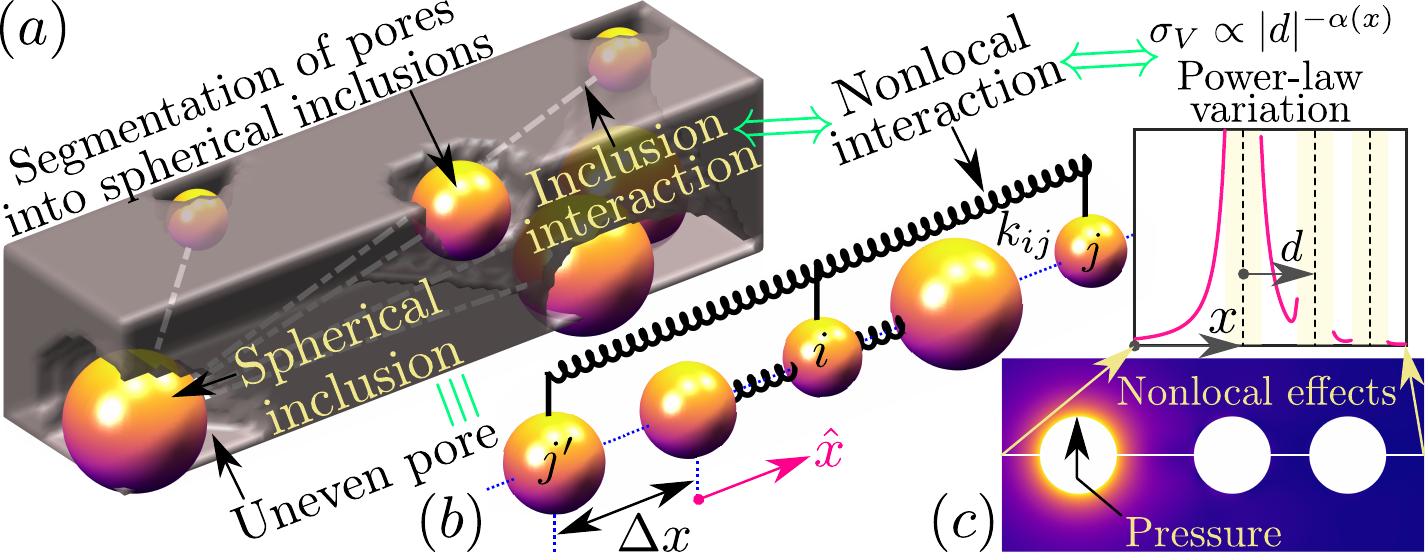}
	\caption{\textbf{Intrinsic variable-order fractional nonlocality in porous solids.} (a) Segmentation of a porous structure into a network of hollow spherical inclusions interacting via long-range forces. Note that an inverted transparency map has been adopted to represent the bulk and the pores in order to facilitate the visualization of the segmentation. (b) Idealized discrete model of long-range interacting spherical inclusions. The discrete nonlocal spring connections capture the redistribution of elastic energy due to the presence of the pores. (c) Power-law spatial variation of the stress in a series of spherical inclusions.}
	\label{fig: PSL}
\end{figure}

\section*{Results and Discussion}
\subsection*{The physics of deformation} 
A porous medium can be approximated as a collection of hollow spherical inclusions dispersed within a bulk solid. While the pores are not necessarily spherical in nature, the total volume of a pore can be effectively segmented into a series of spheres using standard concepts of segmentation \cite{pal1993review} [Fig.~\ref{fig: PSL}(a)]. It follows that a porous structure can be modeled as a network of hollow spherical inclusions dispersed within a solid matrix. Henceforth, we refer to the 'hollow spherical inclusions' as 'spherical inclusions' for the sake of brevity. Further, the spherical inclusions interact with each other via long-range forces [Fig.~\ref{fig: PSL}(a)], hence forming a network of nonlocal interacting particles in the bulk solid. In order to visualize this concept, consider a porous solid with only two inclusions. Any change in the shape of one inclusion (e.g. due to the application of external mechanical forces) will affect also the shape and response of the other inclusion. This example is a direct realization of the principle of \textit{action-at-a-distance} which is at the foundation of atomistic, molecular, and nonlocal continuum theories \cite{eringen1983differential,polizzotto2001nonlocal,zhu2020nonlocal,nair2021nonlocal,russillo2022wave}. Note that the magnitude of the nonlocal force between two inclusions decays monotonically with the increasing (relative) distance following a power-law distribution \cite{golkov2017shape}. This behavior is closely related to the decay of the magnitude of the stress field, which decreases as a power-law function of the radial distance from the spherical inclusion \cite{eshelby1957determination}. Indeed, the presence of a second inclusion within the horizon of influence (that is the region affected by the stress concentration due to the spherical inclusion) will deform under the stress field of the first inclusion and produce, in turn, a stress field around itself that will affect the next inclusion [see Fig.~\ref{fig: PSL}(c)]. It is readily seen how, by following this cascading process, the deformation of the first inclusion can affect the third inclusion (and beyond), hence producing an evident nonlocal effect. Further, the strength of the interaction between inclusions is directly dependent on the concentration of inclusions in a target area as well as on their relative size \cite{ben2015response}. More specifically, the presence of either a larger number of inclusions or of inclusions of larger size will enhance the stress concentration and the corresponding nonlocal effect. It follows that spatial variations in size and distribution of the inclusions will result in variations in the strength of the nonlocal interactions within the network of inclusions. From the above discussion, it clearly emerges that porous media exhibit nonlocal effects characterized by a position-dependent strength strongly influenced by the porous microstructure.

\subsection*{The VO nonlocal elasticity model}
Building on the previous observations of the physical mechanisms taking place in a porous medium, we derive a nonlocal continuum theory capable of capturing the complex nonlocal interactions of the solid. Consider a 1D infinite lattice of inclusions with spatial period $\Delta x$, as shown in Fig.~\ref{fig: PSL}(b) \cite{szajek2019discrete}. The location and displacement of the $n^{th}$ inclusion (where $n\in \mathbb{Z}$) are denoted as $x_n$ and $u_n$, respectively. The nonlocal interactions between the inclusions are modeled as lumped springs. The lumped springs capture both the exchange and the redistribution of elastic energy between the inclusions, and the analogous elastic interactions are indicated as dashed lines within the bulk in Fig.~\ref{fig: PSL}(a). The power-law decay in the interaction force between two inclusions $i$ and $j (\neq i)$ is embedded in the stiffness term characterizing the lumped spring $k_{ij}$:
\begin{equation}
    \label{eq: PL_springs}
    k_{ij}={c_0\psi_0}|x_i-x_j|^{-(1+\alpha(x_i))}
\end{equation}
where $\alpha(x_i)$ captures the position-dependent strength of the nonlocal interactions. To obtain a well-posed and causal theory, it is required that $\alpha(x_i)\in(0,1]$ \cite{patnaik2020generalized}.
Notably, the continuum model with the power-law distribution of order $1+\alpha(x_i)\in(1,2]$ directly reflects the power-law dispersion relations \cite{patnaik2021displacement},  which is widely supported by experimental observations \cite{fellah2000transient,fellah2004ultrasonic,patnaik2020generalized}. Additionally, we also note that strength of long-range forces between spherical inclusions also exhibit power-law distributions with exponents in the interval $(1,2)$ \cite{ben2015response,golkov2017shape}. $c_0$ is a book keeping parameter defined in Equation~(\ref{eq: continuum_limit}). The total internal force on the $i^{th}$ inclusion is given by:
\begin{equation}
    \label{eq: Inclusion_force}
    F(x_i)=\sum_{j=-\infty}^{i-1}k_{ij}(u_i-u_j) + \sum^{j=\infty}_{i+1}k_{ij}(u_j-u_i)
\end{equation}
To facilitate continualization (that is, $\Delta x \rightarrow 0$), $\psi_0$ is defined as $\psi_0 = EAl_*^{\alpha(x_i)-1}\Delta x$, where $E$ and $A$ denote the Young's modulus and cross-sectional area of the equivalent 1D continuum, respectively. 
$l_*$ is a length-scale parameter that ensures dimensional consistency of the formulation [analogous to $l_\mp$ in Equation~(\ref{eq: RC_definition})]. Note that, $\psi_0=EA\Delta x$ for $\alpha(x_i)=1$, indicating that it captures the energy stored in an interaction between adjacent inclusions in a local lattice. Finally, the stress in the 1D nonlocal solid is derived from the continuum limit of the internal force as (see Supplementary Note 1):
\begin{equation}
    \label{eq: continuum_limit}
    \sigma(x) =  \frac{c_0El_*^{\alpha(x)-1}}{\alpha(x)} \int_{-\infty}^{\infty}\frac{1}{|x-x^\prime|^{\alpha(x)}} \left[\frac{\mathrm{d}u(x^\prime)}{\mathrm{d}x^\prime}\right]\mathrm{d}x^\prime
\end{equation}
Assuming $c_0=\alpha(x)/\Gamma(1-\alpha(x))$, where $\Gamma(\cdot)$ denotes the Gamma function, the stress can be expressed in a compact form via VO fractional derivatives:
\begin{equation}
    \label{eq: 1D_stress}
    \sigma(x) = E\underbrace{l_*^{\alpha(x)-1}\left[\;_{-\infty}^C D^{\alpha(x)}_x u - \;_{x}^C D^{\alpha(x)}_\infty u \right]}_{D^{\alpha(x)}_xu \equiv \varepsilon(x)} 
\end{equation}
where $\varepsilon$ denotes the strain in the 1D nonlocal solid, and $\;_{-\infty}^C D^{\alpha(x)}_x u$, $ \;_{x}^C D^{\alpha(x)}_\infty u$ denote the VO left-handed and right-handed Caputo derivatives, respectively. $D^{\alpha(x)}_xu$ denotes the VO Riesz-Caputo derivative.

The above analysis suggests that VO operators are naturally equipped to model porous structures because the differ-integral nature of the VO fractional operator allows capturing the nonlocal interactions, while the spatially varying exponent of the power-law kernel of the operator (that is, $\alpha(x)$) captures the position-dependent strength of the interactions.

In order to leverage the potential of VO operators to model porous structures of practical interest, we develop a 3D VO nonlocal continuum theory by directly extending the 1D strain relations obtained in Equation~(\ref{eq: 1D_stress}):
\begin{equation}
\label{eq: strain}
{\varepsilon}_{ij} = \frac{1}{2} \left[ D^{\alpha(\bm{x})}_{x_j} u_i + D^{\alpha(\bm{x})}_{x_i} u_j + D^{\alpha(\bm{x})}_{x_i} u_m D^{\alpha(\bm{x})}_{x_j} u_m \right]
\end{equation}
where $u_i$ denotes the displacement field along $\hat{x}_i$ direction at a given point $\bm{x}$. Next, the stress tensor is derived from thermodynamic balance laws as:
\begin{equation}
\label{eq: stress}
{\sigma}_{ij} = \lambda\delta_{ij} \varepsilon_{kk} + 2\mu\varepsilon_{ij} - (3\lambda+2\mu) \alpha_\theta\delta_{ij} \theta 
\end{equation}
where $\lambda$ and $\mu$ denote the Lam\'e constants, $\alpha_\theta$ is the coefficient of thermal expansion, and $\theta$ denotes the temperature difference between the solid and the environment. The detailed derivation of the strain and stress tensors can be found in Supplementary Note 2 and Note 5, respectively.

The VO Riesz-Caputo derivative $D^{\alpha(\bm{x})}_{x_j} u_i$ in Equation~(\ref{eq: strain}) is defined on the interval $[x_j^-,x_j^+]$ as:
\begin{equation}
\label{eq: RC_definition}
	\begin{split}
	D^{\alpha(\bm{x})}_{x_j} u_i = \frac{1}{2} \Gamma(2-\alpha(\bm{x})) \Big[ l_{-_j}^{\alpha(\bm{x}) -1}  \;^C_{x^-_j}D^{\alpha(\bm{x})}_{x_j} u_i \\ -  l_{+_j}^{\alpha(\bm{x}) -1}  \;^C_{x_j}D^{\alpha(\bm{x})}_{x^+_j} u_i  \Big]
	\end{split}
\end{equation}
$l_{-_j}$ and $l_{+_j}$ are length scales along the $\hat{x}_j$ direction and ensure both the dimensional consistency and the frame-invariance of the model. Frame-invariance also requires that $l_{-_j} = x_j - x_{-_j}$ and $l_{+_j} = x_{+_j} - x_j$. Physically, the length scales determine the dimensions of the horizon of nonlocality on either side of a point, that is the distance beyond which nonlocal forces cease to act. Detailed discussions on the consistent behavior of the model at material boundaries and interfaces, and the frame-invariance of the model can be found in Supplementary Note 3 and Note 4, respectively.

\subsection*{Overview of the numerical experiments}
\begin{figure}[ht]
	\centering
	\includegraphics[width=\linewidth]{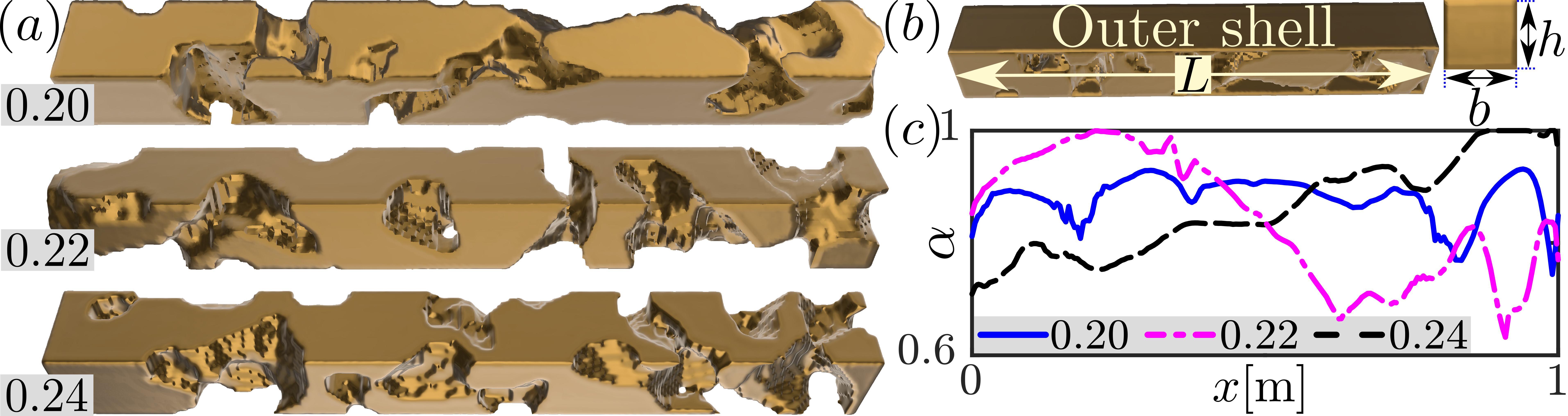}
	\caption{\textbf{Geometry of the different porous beams used for numerical simulations.} (a) Porous cores with porosity $p_0\in\{0.20,0.22,0.24\}$. (b) The beam obtained by enclosing the porous core in a rectangular shell. (c)	Spatial variation of the VO, $\alpha(x)$, across the length of the beam.}
	\label{fig: PA}
\end{figure}

In order to validate and demonstrate the advantages of the VO theory when applied to the modeling of porous structures, we simulate and analyze three configurations of porous beams with porosity $p_0\in\{0.20,0.22,0.24\}$ [Fig.~\ref{fig: PA}(a)]. All beams have length $L=1$m, width $b=0.02$m, height $h=0.02$m, and length scales $l_{-} = l_{+} = 0.2L$ \cite{patnaik2021porous}. The bulk solid has Young's modulus $E_0=30$MPa, Poisson's ratio $\nu=1/3$, and $\alpha_\theta = 2.3 \times 10^{-5} \text{K}^{-1}$. Each porous core is enclosed in a rectangular solid outer shell to simplify the application of external loads and boundary conditions [Fig.~\ref{fig: PA}(b)]. The porous beams were simulated using the nonlinear VO beam model, that was obtained from the 3D theory by employing Euler-Bernoulli assumptions (see Supplementary Note 6). Note that, to apply the VO theory, it is necessary to first determine the VO $\alpha(x)$ that characterizes the underlying microstructural configuration and the resulting nonlocal interactions. In order to determine $\alpha(x)$ [Fig.~\ref{fig: PA}(c)], we formulated an inverse problem by means of deep learning techniques \cite{schuster1997bidirectional}. The details concerning the network architecture and training phase are provided in Supplementary Note 7. With the VO distribution available, the models can be solved numerically using the fractional-order finite element method \cite{sidhardh2020geometrically}, to validate the VO model and assess its performance with respect to the previously identified limitations L1 and L2. Result sets R\# ([R1] Analysis of computational performance, and [R2] Pore configuration, nonlocal effects, and role of $\alpha(x)$) map directly to the limitations L\#.

\begin{figure*}[ht]
	\centering
	\includegraphics[width=\textwidth]{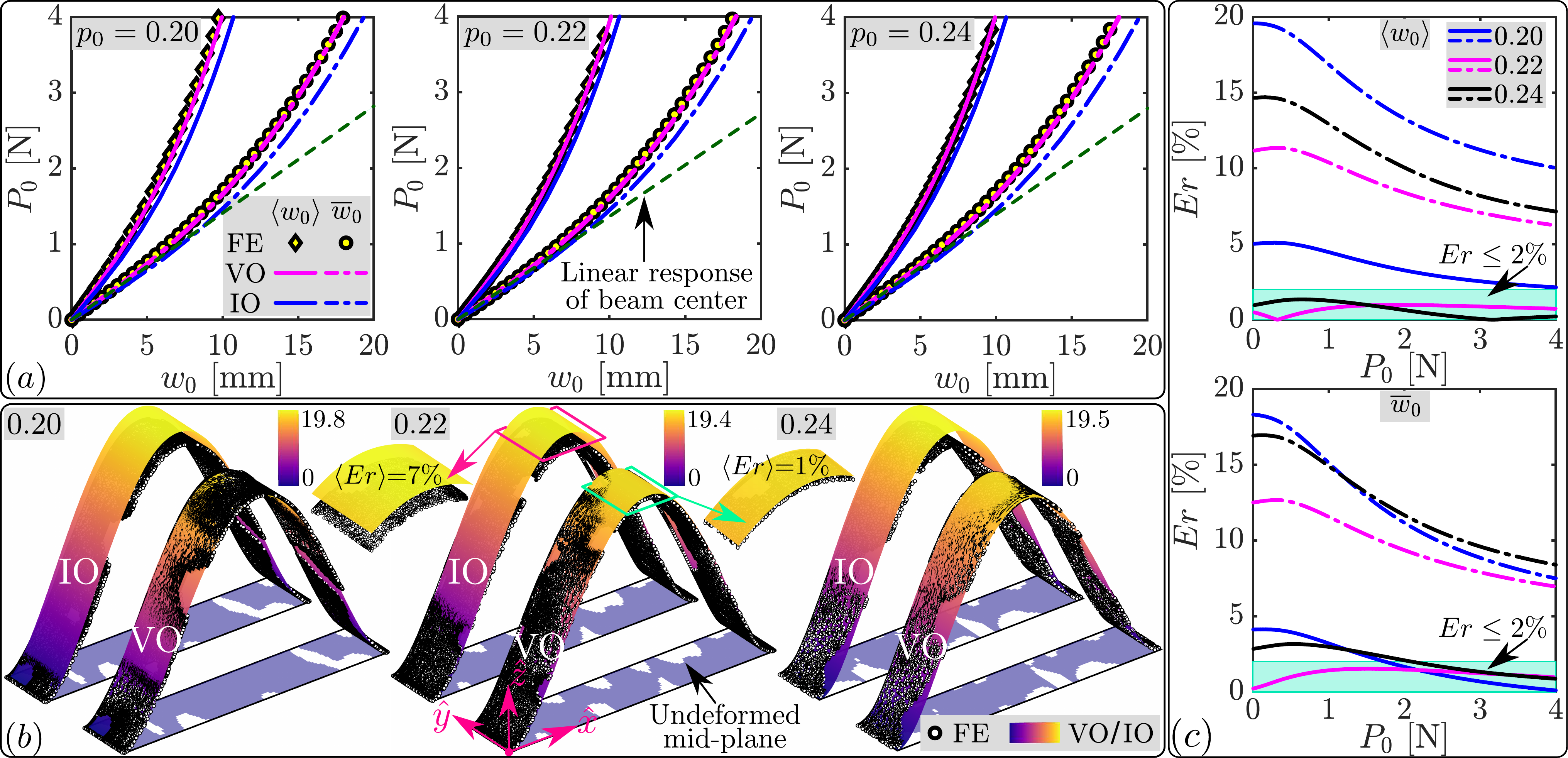}
	\caption{\textbf{Nonlinear static response of the porous beams.} (a) All the beams are clamped at both ends and subject to a UDTL $P_0$. Nonlinear response obtained via three approaches (3D FE, VO, and IO) are compared in terms of the average transverse displacement $(\langle w_0 \rangle)$ and the transverse displacement of the center of the beam $(\overline{w}_0)$. The linear response of the center of the beam (dashed dark-green line) is provided to highlight the impact of geometric nonlinearity on the response of the beam. The linear response is simulated by ignoring the nonlinear term $\left[ D^{\alpha(\bm{x})}_{x_i} u_m D^{\alpha(\bm{x})}_{x_j} u_m \right]$ within the definition of 3D VO strain tensor in Equation~(\ref{eq: strain}). (b) Deformed mid-planes obtained from the VO and IO models are compared against FE results for $P_0=4$N. Spatial variation of the transverse displacement is illustrated on the deformed planes via color maps. The insets for $p_0=0.22$ highlight the error in the region $x\times y =$ $[0.4,0.6]L\times[0,b]$. (c) Relative error in the predictions of the VO (solid line) and IO (dashed line) models with respect to the FE results (considered as ground truth).}
	\label{fig: Static_response}
\end{figure*}

\begin{figure*}[h!]
	\centering
	\includegraphics[width=\textwidth]{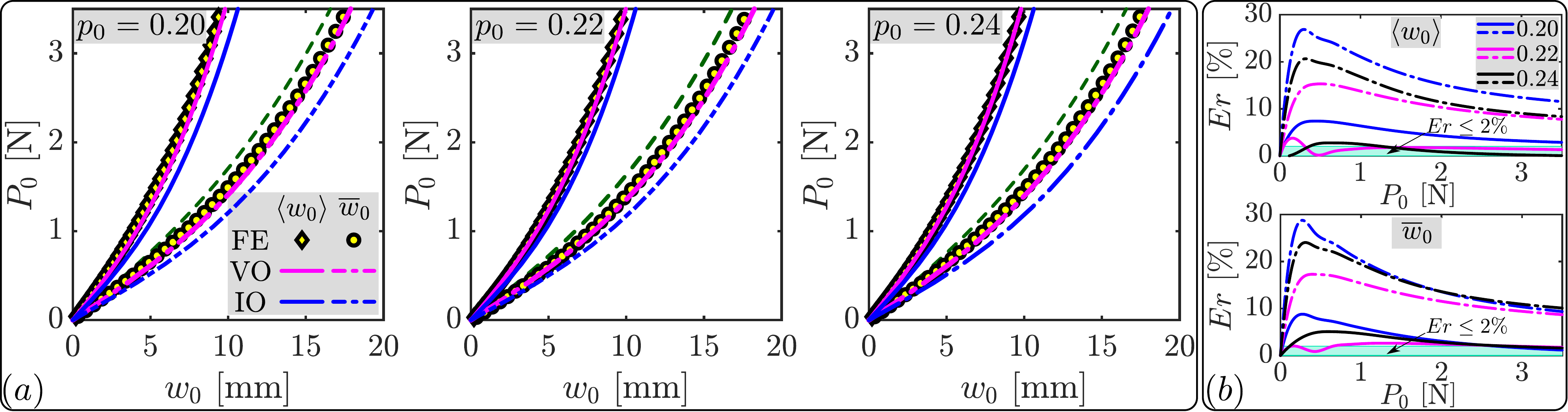}
	\caption{\textbf{Nonlinear thermoelastic response of the porous beams.} (a) All the beams are subjected to a UDTL $P_0$, an isothermal load $\theta=100K$, and are clamped at both ends. The nonlinear response obtained via the three different methods are compared in terms of the average transverse displacement $(\langle w_0 \rangle)$ and the transverse displacement of the geometric center $(\overline{w}_0)$. Response of the center of the beam in the absence of thermal loads (dashed dark-green line) is provided to highlight the consistent softening. (b) Relative error in the predictions of the VO (solid line) and IO (dashed line) models with respect to the FE results (considered as the ground truth).} 
	\label{fig: Thermoelastic_response}
\end{figure*}

The numerical accuracy of the VO model can be assessed by considering the nonlinear static response of the porous beams. All configurations are assumed clamped at both ends and subject to a uniformly distributed transverse load (UDTL) of magnitude $P_0$.
To provide meaningful comparisons with established methods widely adopted in literature, the beams' response was also simulated by using the \textit{3D finite element (FE) approach} and the \textit{rule of mixtures approach} \cite{madsen2003physical,anirudh2019comprehensive}. The 3D FE approach, implemented in this study via the commercial FE software \texttt{COMSOL Multiphysics}, is chosen because it can fully resolve the porous geometry of the beam and hence it provides a reliable reference solution for performance assessment. The rule of mixtures approach is an integer-order (IO) homogenization approach that converts the initial porous beam into an anisotropic (non-porous) beam with a spatially varying modulus of elasticity $E(x) = [(1-p_0(x)]E_0$, where $p_0(x)$ denotes the average porosity of planes perpendicular to the mid-plane of the beam (see Supplementary Note 8). The IO approach allows highlighting the remarkable effect on accuracy that VO mechanics offers over classical IO mechanics.

The nonlinear static response of the porous beams are presented in Fig.~\ref{fig: Static_response}(a) in terms of the average transverse displacement ($\langle w_0 \rangle$) and the transverse displacement of the geometric center [$\bm{x}(L/2,0,0)$; see Fig.~S2] of the beam ($\overline{w}_0$).
Note that $\langle w_0 \rangle$ presents a global comparison of the three approaches, while $\overline{w}_0$ allows a point-wise comparison. To further explore the point-wise comparison, the deformed shapes of the mid-planes for $P_0=4$N are presented in Fig.~\ref{fig: Static_response}(b).
Additionally, for a more complete validation, we also simulated the thermoelastic response of the porous beams when subjected to a steady state uniform temperature $\theta=100$K, in addition to the UDTL $P_0$. The results of the thermoelastic study are presented in Fig.~\ref{fig: Thermoelastic_response}.

\begin{figure}[ht!]
	\centering
	\includegraphics[width=0.85\linewidth]{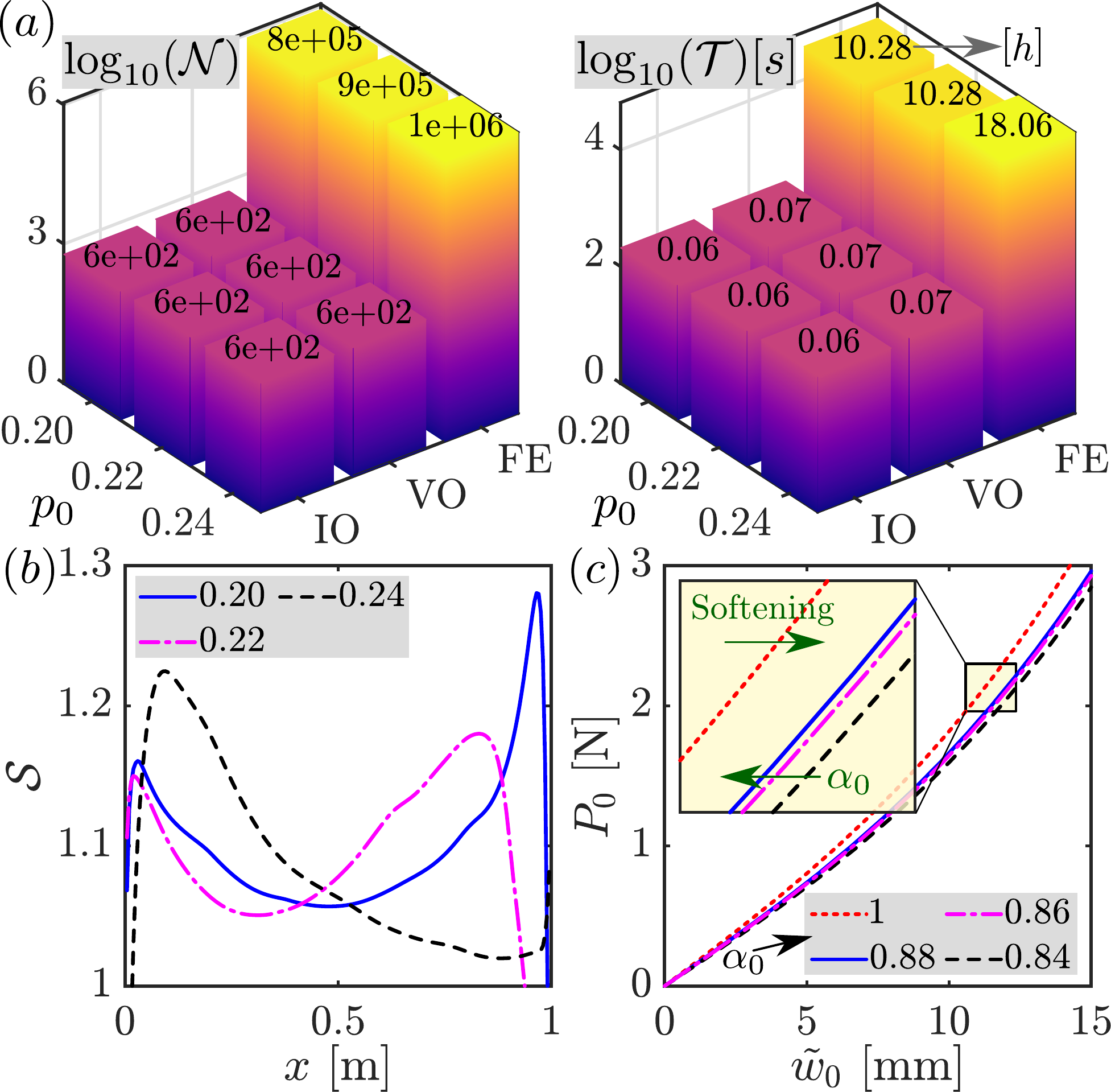}
	\caption{\textbf{Computational performance and some theoretical implications of the VO model.} (a) Performance of the 3D FE, VO, and IO approaches are compared in terms of the degrees of freedom $(\mathcal{N})$ and simulation times $(\mathcal{T})$. For a better understanding, values of $\mathcal{N}$ and $\mathcal{T}$ (in hours) are provided on the top of each bar. (b-c) Impact of nonlocal effects: (b) position-dependent softening of the porous beams across the symmetric axis. (c) Consistent softening of the porous beams with an increase in the degree of nonlocality (that is, a decrease in the average VO $\alpha_0$). The dark-green arrows depict the direction in which either the average VO $\alpha_0$ or the degree of softening increase.}
	\label{fig: PPS}
\end{figure}

\subsection*{Analysis of computational performance}
The total number of degrees of freedom and the average run time (for three repeated simulations) are presented in Fig.~\ref{fig: PPS} for each method. The number of degrees of freedom was chosen so to guarantee proper resolution of the displacement profile. This criterion was defined by setting an arbitrary threshold of $1\%$ on the magnitude of the difference of the displacement field between successive spatial refinements. The VO and IO models were simulated on a personal computer equipped with an Intel(R) Core(TM) i7-1065G7 processor and $8$ GB RAM, and required less than 2 minutes run time. The FE models were run on a cluster with $2\times$AMD EPYC $7662$ processors with $20$ cores and $256$ GB RAM, and required at least 10 hours run time\footnote{\url{https://www.rcac.purdue.edu/compute/bell}}. While the two platforms are different, the cluster is significantly more powerful than the personal computer, hence making the comparison of the computational performance even more conservative. It follows that, based on the above results, the VO approach delivers an unparalleled computational efficiency (compared with the 3D FE approach) while simultaneously maintaining same level of accuracy. In this regard, note that the VO approach can achieve superior accuracy when compared to the rule of mixtures approach. This should not be surprising since the VO model fully captures the position-dependent nonlocal interactions between the pores (as we will demonstrate further in R2), while the the rule of mixtures model is a local approach. 

\subsection*{Pore configuration, nonlocal effects, and role of $\alpha(x)$} 
The different results (in R1) presented for the numerical validation and analysis of the performance, suggest that the VO approach provides a physically meaningful parameter (that is, the VO $\alpha(x)$) which accurately captures the characteristics of the underlying porous microstructure, hence enabling an efficient continuum level formulation endowed with detailed microscale information. Indeed, as motivated via the lattice model, the VO $\alpha(x)$ directly links the pores' configuration to the strength of the nonlocal effects. This latter concept is further supported, here below, by the results of the VO model that consistently captures the impact of the nonlocal effects on the anomalous softening behavior of the porous beams.

At the macro scales, the most direct impact of the presence of nonlocal effects is observed as structural softening, wherein the strength of the nonlocal effects determine the degree of softening \cite{patnaik2021displacement}. In this regard, for porous beams, the position-dependent strength of the nonlocal interactions (due to the spatially-varying porous microstructure) is expected to result in a spatial variation in the degree of softening, when compared to an isotropic non-porous beam ($p_0=0$). Indeed, the VO model successfully captures the position-dependent softening of the porous beam, as evident from the variation of $\mathcal{S}=w_0(p_0\neq 0)/w_0(p_0=0)$ across the length of the beam [Fig.~\ref{fig: PPS}(b)]. The parameter $\mathcal{S}$ captures the increase in displacement at a given point on the porous beam, when compared to an isotropic non-porous beam, on account of the nonlocal interactions resulting from the porous microstructure. This result establishes clearly that the simplifying assumption of spatially-uniform size-effects in classical nonlocal and micro-mechanical models introduces severe inaccuracies.

Further, an analysis of the maximum transverse displacement ($\tilde{w}_0$) [Fig.~\ref{fig: PPS}(c)] shows that an increase in the average porosity, for a fixed $P_0$, does not result in a consistent softening of the porous beam (that is, an increase in $\tilde{w}_0$). This observation is in contrast to the expected behavior based on a classical IO model. However, an increase in the strength of nonlocality (e.g. via a decrease in the average VO [$\alpha_0=\langle \alpha(x) \rangle$; $\alpha_0$ for beams with $p_0=\{0,0.20,0.22,0.24\}$ are $ \alpha_0 = \{1,0.88,0.84,0.86\}$]) correlates in a consistent manner with the softening phenomenon\footnote{In constant-order fractional kinematic approaches to nonlocal elasticity, a decrease in the order results in an increase in the strength of the nonlocal effects (degree of nonlocality) and in a consistent softening behavior \cite{patnaik2021displacement}.}. Thus, it emerges that at the continuum level the average nonlocal strength $\alpha_0$, and not the average porosity $p_0$, provides a much more appropriate and physically consistent representation of the porous microstructure. On a more theoretical note, we emphasize that the softening behavior observed in the porous beams (in both the nonlinear elastic and thermoelastic response) was captured consistently (that is, without any display of paradoxical effects \cite{challamel2008small,batra159misuse}) by the VO theory. This result is possible thanks to the underlying foundation provided by the displacement-driven formulation [based on differ-integral kinematic relations, see Equation~(\ref{eq: strain})], which is intrinsically well-posed and thermodynamically consistent \cite{patnaik2021displacement}.

For completeness, we also highlight that the presence of nonlocal effects does not always lead to softening behavior at the continuum level. Particularly, in the case of nonlocal solids with multiscale effects, an increase in the strength of nonlocality could potentially lead to either stiffening or softening behavior \cite{patnaik2020towards,ding2021multiscale}. In this case, a different approach should be used, for example, distributed-order fractional models \cite{ding2021multiscale} or fractional-order strain gradients in the nonlocal constitutive formulation \cite{patnaik2020towards}. Nevertheless, in the present study, the radii of the pores distributed within the bulk are within the same scale (or, order) when compared to the scale of the beam and hence, we do not account for multiscale effects. As also verified numerically (based on the fully resolved geometry via 3D finite element analysis), the porous beams exhibit a consistent softening behavior indicating the absence of any multiscale effects.

\subsection*{Outlook}
In summary, this study has provided two key contributions to the physical understanding and modeling of porous media. First, it was shown that porous solids exhibit position-dependent nonlocal effects that cannot be neglected if accurate predictions are sought. Second, a VO nonlocal continuum theory capable of capturing these complex nonlocal effects was developed and numerically tested. Results highlighted that VO mechanics offers a uniquely powerful approach to develop efficient continuum models endowed with fine level details capturing the underlying microstructure. The VO approach provides a rigorous methodology to develop physically-consistent reduced-order models of multiscale systems with an accuracy comparable with fully resolved 3D models. While the results were presented in the context of porous materials, it is expected that the VO framework could be extended to a variety of applications characterized by multiscale features including, but not limited to, composites, architectured materials, seismology, biotechnology, and much more.

\section*{Methods}
The key methods adopted in deriving the results include: [M1] the use of VO fractional calculus to develop the VO nonlocal elasticity model presented in the results; and [M2] the use of a deep learning method to obtain the VO laws that characterize the porous solids. These methods have been described in detail in the supplementary information document. The method M1 has been divided into specific analytical derivations which are presented in the Supplementary Notes 1 - 6 and summarized here below:
\begin{itemize}[leftmargin=*]
    \item Supplementary Note 1 presents the analytical continualization of the discrete lattice to obtain the stress and strain fields in the 1D continuum, that is essentially the derivation of Equation~(\ref{eq: Inclusion_force}) from Equation~(\ref{eq: 1D_stress}).
    
    \item Supplementary Note 2 presents the derivation of the 3D VO nonlocal strain tensor.
    
    \item Supplementary Note 3 presents the analytical proof of the consistent behavior of the VO formulation at material boundaries and interfaces.
    
    \item Supplementary Note 4 presents the frame-invariance of the VO nonlocal continuum model.
    
    \item Supplementary Note 5 presents the derivation of the stress tensor and the proof of the thermodynamic consistency of the VO nonlocal continuum model.
    
    \item Supplementary Note 6 presents the derivation of a geometrically nonlinear model for nonlocal slender beams from the 3D VO nonlocal continuum model.
    
    \item Supplementary Note 7 presents the deep learning framework for the identification of the VO that characterizes the response of different nonlocal beams.
\end{itemize}
The method M2 is presented in Supplementary Note 7.

\section*{Data Availability}
All the necessary data and information required to reproduce the results are available in the paper. Additional data in the form of equations supporting this article are provided in the supplementary information.

\section*{Code Availability}
The files of the numerical simulation data corresponding to the results in the paper will also be shared. The actual physical code may be subject to restrictions from the sponsors.

\section*{Acknowledgements}
The authors gratefully acknowledge the financial support of the National Science Foundation under grants MOMS \#1761423, DCSD \#1825837, and CAREER \#1621909, and the Defense Advanced Research Project Agency under grant \#D19AP00052. S.P. acknowledges the support of the School of Mechanical Engineering, Purdue University, through the Hugh W. and Edna M. Donnan Fellowship. Any opinions, findings, and conclusions or recommendations expressed in this material are those of the author(s) and do not necessarily reflect the views of the National Science Foundation. The content and information presented in this manuscript do not necessarily reflect the position or the policy of the government. The material is approved for public release; distribution is unlimited.

\section*{Author Contributions} 
S.P. and F.S. conceptualized and designed the research; S.P. performed the research; S.P., M.J., W.D., and F.S. contributed new analytic tools; S.P. and F.S. analyzed data; and S.P., M.J., W.D., and F.S. wrote the paper.

\section*{Competing Interests}
The authors declare that there are no competing interests.


\begin{thebibliography}{10}
\expandafter\ifx\csname url\endcsname\relax
  \def\url#1{\texttt{#1}}\fi
\expandafter\ifx\csname urlprefix\endcsname\relax\def\urlprefix{URL }\fi
\providecommand{\bibinfo}[2]{#2}
\providecommand{\eprint}[2][]{\url{#2}}

\bibitem{day2021evolution}
\bibinfo{author}{Day, G.~S.}, \bibinfo{author}{Drake, H.~F.},
  \bibinfo{author}{Zhou, H.-C.} \& \bibinfo{author}{Ryder, M.~R.}
\newblock \bibinfo{title}{Evolution of porous materials from ancient remedies
  to modern frameworks}.
\newblock \emph{\bibinfo{journal}{Commun. Chem.}} \textbf{\bibinfo{volume}{4}},
  \bibinfo{pages}{1--4} (\bibinfo{year}{2021}).

\bibitem{zou2017porous}
\bibinfo{author}{Zou, L.} \emph{et~al.}
\newblock \bibinfo{title}{Porous organic polymers for post-combustion carbon
  capture}.
\newblock \emph{\bibinfo{journal}{Adv. Mater.}} \textbf{\bibinfo{volume}{29}},
  \bibinfo{pages}{1700229} (\bibinfo{year}{2017}).

\bibitem{krishna2008engineered}
\bibinfo{author}{Krishna, B.~V.}, \bibinfo{author}{Xue, W.},
  \bibinfo{author}{Bose, S.} \& \bibinfo{author}{Bandyopadhyay, A.}
\newblock \bibinfo{title}{Engineered porous metals for implants}.
\newblock \emph{\bibinfo{journal}{JOM}} \textbf{\bibinfo{volume}{60}},
  \bibinfo{pages}{45--48} (\bibinfo{year}{2008}).

\bibitem{fellah2004ultrasonic}
\bibinfo{author}{Fellah, Z. E.~A.}, \bibinfo{author}{Chapelon, J.~Y.},
  \bibinfo{author}{Berger, S.}, \bibinfo{author}{Lauriks, W.} \&
  \bibinfo{author}{Depollier, C.}
\newblock \bibinfo{title}{Ultrasonic wave propagation in human cancellous bone:
  Application of {B}iot theory}.
\newblock \emph{\bibinfo{journal}{J. Acoust. Soc. Am.}}
  \textbf{\bibinfo{volume}{116}}, \bibinfo{pages}{61--73}
  (\bibinfo{year}{2004}).

\bibitem{mannan2017correlations}
\bibinfo{author}{Mannan, S.}, \bibinfo{author}{Paul~Knox, J.} \&
  \bibinfo{author}{Basu, S.}
\newblock \bibinfo{title}{Correlations between axial stiffness and
  microstructure of a species of bamboo}.
\newblock \emph{\bibinfo{journal}{R. Soc. Open Sci.}}
  \textbf{\bibinfo{volume}{4}}, \bibinfo{pages}{160412} (\bibinfo{year}{2017}).

\bibitem{mannan2018stiffness}
\bibinfo{author}{Mannan, S.}, \bibinfo{author}{Parameswaran, V.} \&
  \bibinfo{author}{Basu, S.}
\newblock \bibinfo{title}{Stiffness and toughness gradation of bamboo from a
  damage tolerance perspective}.
\newblock \emph{\bibinfo{journal}{Int. J. Solids Struct.}}
  \textbf{\bibinfo{volume}{143}}, \bibinfo{pages}{274--286}
  (\bibinfo{year}{2018}).

\bibitem{zhao2014bioinspired}
\bibinfo{author}{Zhao, N.} \emph{et~al.}
\newblock \bibinfo{title}{Bioinspired materials: from low to high dimensional
  structure}.
\newblock \emph{\bibinfo{journal}{Adv. Mater.}} \textbf{\bibinfo{volume}{26}},
  \bibinfo{pages}{6994--7017} (\bibinfo{year}{2014}).

\bibitem{gibson2000mechanical}
\bibinfo{author}{Gibson, L.}
\newblock \bibinfo{title}{Mechanical behavior of metallic foams}.
\newblock \emph{\bibinfo{journal}{Annu. Rev. Mater. Sci.}}
  \textbf{\bibinfo{volume}{30}}, \bibinfo{pages}{191--227}
  (\bibinfo{year}{2000}).

\bibitem{chouksey2019computational}
\bibinfo{author}{Chouksey, M.}, \bibinfo{author}{Keralavarma, S.~M.} \&
  \bibinfo{author}{Basu, S.}
\newblock \bibinfo{title}{Computational investigation into the role of
  localisation on yield of a porous ductile solid}.
\newblock \emph{\bibinfo{journal}{J. Mech. Phys. Solids}}
  \textbf{\bibinfo{volume}{130}}, \bibinfo{pages}{141--164}
  (\bibinfo{year}{2019}).

\bibitem{laubie2017stress}
\bibinfo{author}{Laubie, H.}, \bibinfo{author}{Radjai, F.},
  \bibinfo{author}{Pellenq, R.} \& \bibinfo{author}{Ulm, F.-J.}
\newblock \bibinfo{title}{Stress transmission and failure in disordered porous
  media}.
\newblock \emph{\bibinfo{journal}{Phys. Rev. Lett.}}
  \textbf{\bibinfo{volume}{119}}, \bibinfo{pages}{075501}
  (\bibinfo{year}{2017}).

\bibitem{fellah2000transient}
\bibinfo{author}{Fellah, Z.} \& \bibinfo{author}{Depollier, C.}
\newblock \bibinfo{title}{Transient acoustic wave propagation in rigid porous
  media: A time-domain approach}.
\newblock \emph{\bibinfo{journal}{J. Acoust. Soc. Am.}}
  \textbf{\bibinfo{volume}{107}}, \bibinfo{pages}{683--688}
  (\bibinfo{year}{2000}).

\bibitem{grosman2008influence}
\bibinfo{author}{Grosman, A.} \& \bibinfo{author}{Ortega, C.}
\newblock \bibinfo{title}{Influence of elastic deformation of porous materials
  in adsorption-desorption process: A thermodynamic approach}.
\newblock \emph{\bibinfo{journal}{Phys. Rev. B}} \textbf{\bibinfo{volume}{78}},
  \bibinfo{pages}{085433} (\bibinfo{year}{2008}).

\bibitem{hellstrom2010mechanisms}
\bibinfo{author}{Hellstr{\"o}m, J. G.~I.}, \bibinfo{author}{Frishfelds, V.} \&
  \bibinfo{author}{Lundstr{\"o}m, T.}
\newblock \bibinfo{title}{Mechanisms of flow-induced deformation of porous
  media}.
\newblock \emph{\bibinfo{journal}{J. Fluid Mech.}}
  \textbf{\bibinfo{volume}{664}}, \bibinfo{pages}{220--237}
  (\bibinfo{year}{2010}).

\bibitem{patnaik2021porous}
\bibinfo{author}{Patnaik, S.}, \bibinfo{author}{Jokar, M.} \&
  \bibinfo{author}{Semperlotti, F.}
\newblock \bibinfo{title}{Variable-order approach to nonlocal elasticity:
  Theoretical formulation, order identification via deep learning, and
  applications}.
\newblock \emph{\bibinfo{journal}{Comput. Mech.}}
  \textbf{\bibinfo{volume}{69}}, \bibinfo{pages}{267–298}
  (\bibinfo{year}{2021}).

\bibitem{madsen2003physical}
\bibinfo{author}{Madsen, B.} \& \bibinfo{author}{Lilholt, H.}
\newblock \bibinfo{title}{Physical and mechanical properties of unidirectional
  plant fibre composites—an evaluation of the influence of porosity}.
\newblock \emph{\bibinfo{journal}{Compos. Sci. Technol.}}
  \textbf{\bibinfo{volume}{63}}, \bibinfo{pages}{1265--1272}
  (\bibinfo{year}{2003}).

\bibitem{anirudh2019comprehensive}
\bibinfo{author}{Anirudh, B.}, \bibinfo{author}{Ganapathi, M.},
  \bibinfo{author}{Anant, C.} \& \bibinfo{author}{Polit, O.}
\newblock \bibinfo{title}{A comprehensive analysis of porous
  graphene-reinforced curved beams by finite element approach using
  higher-order structural theory: Bending, vibration and buckling}.
\newblock \emph{\bibinfo{journal}{Compos. Struct.}}
  \textbf{\bibinfo{volume}{222}}, \bibinfo{pages}{110899}
  (\bibinfo{year}{2019}).

\bibitem{batra159misuse}
\bibinfo{author}{Batra, R.}
\newblock \bibinfo{title}{Misuse of {E}ringen's nonlocal elasticity theory for
  functionally graded materials}.
\newblock \emph{\bibinfo{journal}{Int. J. Eng. Sci.}}
  \textbf{\bibinfo{volume}{159}}, \bibinfo{pages}{103425}
  (\bibinfo{year}{2021}).

\bibitem{goransson2008tailored}
\bibinfo{author}{G{\"o}ransson, P.}
\newblock \bibinfo{title}{Tailored acoustic and vibrational damping in porous
  solids--engineering performance in aerospace applications}.
\newblock \emph{\bibinfo{journal}{Aerosp. Sci. Technol.}}
  \textbf{\bibinfo{volume}{12}}, \bibinfo{pages}{26--41}
  (\bibinfo{year}{2008}).

\bibitem{xue2019structural}
\bibinfo{author}{Xue, Y.}, \bibinfo{author}{Bolton, J.~S.},
  \bibinfo{author}{Herdtle, T.}, \bibinfo{author}{Lee, S.} \&
  \bibinfo{author}{Gerdes, R.~W.}
\newblock \bibinfo{title}{Structural damping by lightweight poro-elastic
  media}.
\newblock \emph{\bibinfo{journal}{J. Sound Vib.}}
  \textbf{\bibinfo{volume}{459}}, \bibinfo{pages}{114866}
  (\bibinfo{year}{2019}).

\bibitem{zhao2016understanding}
\bibinfo{author}{Zhao, K.} \& \bibinfo{author}{Cui, Y.}
\newblock \bibinfo{title}{Understanding the role of mechanics in energy
  materials: A perspective}.
\newblock \emph{\bibinfo{journal}{Extreme Mech. Lett.}}
  \textbf{\bibinfo{volume}{9}}, \bibinfo{pages}{347--352}
  (\bibinfo{year}{2016}).

\bibitem{li2022dynamics}
\bibinfo{author}{Li, J.} \emph{et~al.}
\newblock \bibinfo{title}{Dynamics of particle network in composite battery
  cathodes}.
\newblock \emph{\bibinfo{journal}{Science}} \textbf{\bibinfo{volume}{376}},
  \bibinfo{pages}{517--521} (\bibinfo{year}{2022}).

\bibitem{samko1993integration}
\bibinfo{author}{Samko, S.~G.} \& \bibinfo{author}{Ross, B.}
\newblock \bibinfo{title}{Integration and differentiation to a variable
  fractional order}.
\newblock \emph{\bibinfo{journal}{Integral Transforms Spec. Funct.}}
  \textbf{\bibinfo{volume}{1}}, \bibinfo{pages}{277--300}
  (\bibinfo{year}{1993}).

\bibitem{chechkin2005fractional}
\bibinfo{author}{Chechkin, A.~V.}, \bibinfo{author}{Gorenflo, R.} \&
  \bibinfo{author}{Sokolov, I.~M.}
\newblock \bibinfo{title}{Fractional diffusion in inhomogeneous media}.
\newblock \emph{\bibinfo{journal}{J. Phys. A}} \textbf{\bibinfo{volume}{38}},
  \bibinfo{pages}{L679} (\bibinfo{year}{2005}).

\bibitem{sokolov2006field}
\bibinfo{author}{Sokolov, I.~M.} \& \bibinfo{author}{Klafter, J.}
\newblock \bibinfo{title}{Field-induced dispersion in subdiffusion}.
\newblock \emph{\bibinfo{journal}{Phys. Rev. Lett.}}
  \textbf{\bibinfo{volume}{97}}, \bibinfo{pages}{140602}
  (\bibinfo{year}{2006}).

\bibitem{di2020novel}
\bibinfo{author}{Di~Paola, M.}, \bibinfo{author}{Alotta, G.},
  \bibinfo{author}{Burlon, A.} \& \bibinfo{author}{Failla, G.}
\newblock \bibinfo{title}{A novel approach to nonlinear variable-order
  fractional viscoelasticity}.
\newblock \emph{\bibinfo{journal}{Phil. Trans. R. Soc. A}}
  \textbf{\bibinfo{volume}{378}}, \bibinfo{pages}{20190296}
  (\bibinfo{year}{2020}).

\bibitem{suzuki2021fractional}
\bibinfo{author}{Suzuki, J.}, \bibinfo{author}{Gulian, M.},
  \bibinfo{author}{Zayernouri, M.} \& \bibinfo{author}{D'Elia, M.}
\newblock \bibinfo{title}{Fractional modeling in action: A survey of nonlocal
  models for subsurface transport, turbulent flows, and anomalous materials}.
\newblock {\bibinfo{journal}{Preprint at \url{https://arxiv.org/abs/2110.11531}}}
  (\bibinfo{year}{2021}).

\bibitem{patnaik2020review}
\bibinfo{author}{Patnaik, S.}, \bibinfo{author}{Hollkamp, J.~P.} \&
  \bibinfo{author}{Semperlotti, F.}
\newblock \bibinfo{title}{Applications of variable-order fractional operators:
  a review}.
\newblock \emph{\bibinfo{journal}{Proc. R. Soc. A}}
  \textbf{\bibinfo{volume}{476}}, \bibinfo{pages}{20190498}
  (\bibinfo{year}{2020}).

\bibitem{west2016fractional}
\bibinfo{author}{West, B.~J.}
\newblock \emph{\bibinfo{title}{Fractional calculus view of complexity:
  tomorrow's science}}. 1st edn (\bibinfo{publisher}{CRC Press}, \bibinfo{year}{2016}).

\bibitem{patnaik2020generalized}
\bibinfo{author}{Patnaik, S.} \& \bibinfo{author}{Semperlotti, F.}
\newblock \bibinfo{title}{A generalized fractional-order elastodynamic theory
  for non-local attenuating media}.
\newblock \emph{\bibinfo{journal}{Proc. R. Soc. A}}
  \textbf{\bibinfo{volume}{476}}, \bibinfo{pages}{20200200}
  (\bibinfo{year}{2020}).

\bibitem{pal1993review}
\bibinfo{author}{Pal, N.~R.} \& \bibinfo{author}{Pal, S.~K.}
\newblock \bibinfo{title}{A review on image segmentation techniques}.
\newblock \emph{\bibinfo{journal}{Pattern Recognit.}}
  \textbf{\bibinfo{volume}{26}}, \bibinfo{pages}{1277--1294}
  (\bibinfo{year}{1993}).

\bibitem{eringen1983differential}
\bibinfo{author}{Eringen, A.~C.}
\newblock \bibinfo{title}{On differential equations of nonlocal elasticity and
  solutions of screw dislocation and surface waves}.
\newblock \emph{\bibinfo{journal}{J. Appl. Phys.}}
  \textbf{\bibinfo{volume}{54}}, \bibinfo{pages}{4703--4710}
  (\bibinfo{year}{1983}).

\bibitem{polizzotto2001nonlocal}
\bibinfo{author}{Polizzotto, C.}
\newblock \bibinfo{title}{Nonlocal elasticity and related variational
  principles}.
\newblock \emph{\bibinfo{journal}{Int. J. Solids Struct.}}
  \textbf{\bibinfo{volume}{38}}, \bibinfo{pages}{7359--7380}
  (\bibinfo{year}{2001}).

\bibitem{zhu2020nonlocal}
\bibinfo{author}{Zhu, H.}, \bibinfo{author}{Patnaik, S.},
  \bibinfo{author}{Walsh, T.~F.}, \bibinfo{author}{Jared, B.~H.} \&
  \bibinfo{author}{Semperlotti, F.}
\newblock \bibinfo{title}{Nonlocal elastic metasurfaces: Enabling broadband
  wave control via intentional nonlocality}.
\newblock \emph{\bibinfo{journal}{Proc. Natl. Acad. Sci. U.S.A.}}
  \textbf{\bibinfo{volume}{117}}, \bibinfo{pages}{26099--26108}
  (\bibinfo{year}{2020}).

\bibitem{nair2021nonlocal}
\bibinfo{author}{Nair, S.}, \bibinfo{author}{Jokar, M.} \&
  \bibinfo{author}{Semperlotti, F.}
\newblock \bibinfo{title}{Nonlocal acoustic black hole metastructures:
  {A}chieving broadband and low frequency passive vibration attenuation}.
\newblock \emph{\bibinfo{journal}{Mech. Syst. Signal Process.}}
  \textbf{\bibinfo{volume}{169}}, \bibinfo{pages}{108716}
  (\bibinfo{year}{2022}).

\bibitem{russillo2022wave}
\bibinfo{author}{Russillo, A.~F.} \& \bibinfo{author}{Failla, G.}
\newblock \bibinfo{title}{Wave propagation in stress-driven nonlocal {R}ayleigh
  beam lattices}.
\newblock \emph{\bibinfo{journal}{Int. J. Mech. Sci.}}
  \textbf{\bibinfo{volume}{215}}, \bibinfo{pages}{106901}
  (\bibinfo{year}{2022}).

\bibitem{golkov2017shape}
\bibinfo{author}{Golkov, R.} \& \bibinfo{author}{Shokef, Y.}
\newblock \bibinfo{title}{Shape regulation generates elastic interaction
  between living cells}.
\newblock \emph{\bibinfo{journal}{New J. Phys.}} \textbf{\bibinfo{volume}{19}},
  \bibinfo{pages}{063011} (\bibinfo{year}{2017}).

\bibitem{eshelby1957determination}
\bibinfo{author}{Eshelby, J.~D.}
\newblock \bibinfo{title}{The determination of the elastic field of an
  ellipsoidal inclusion, and related problems}.
\newblock \emph{\bibinfo{journal}{Proc. R. Soc. A}}
  \textbf{\bibinfo{volume}{241}}, \bibinfo{pages}{376--396}
  (\bibinfo{year}{1957}).

\bibitem{ben2015response}
\bibinfo{author}{Ben-Yaakov, D.}, \bibinfo{author}{Golkov, R.},
  \bibinfo{author}{Shokef, Y.} \& \bibinfo{author}{Safran, S.~A.}
\newblock \bibinfo{title}{Response of adherent cells to mechanical
  perturbations of the surrounding matrix}.
\newblock \emph{\bibinfo{journal}{Soft Matter}} \textbf{\bibinfo{volume}{11}},
  \bibinfo{pages}{1412--1424} (\bibinfo{year}{2015}).

\bibitem{szajek2019discrete}
\bibinfo{author}{Szajek, K.} \& \bibinfo{author}{Sumelka, W.}
\newblock \bibinfo{title}{Discrete mass-spring structure identification in
  nonlocal continuum space-fractional model}.
\newblock \emph{\bibinfo{journal}{Eur. Phys. J. Plus}}
  \textbf{\bibinfo{volume}{134}}, \bibinfo{pages}{1--19}
  (\bibinfo{year}{2019}).

\bibitem{patnaik2021displacement}
\bibinfo{author}{Patnaik, S.}, \bibinfo{author}{Sidhardh, S.} \&
  \bibinfo{author}{Semperlotti, F.}
\newblock \bibinfo{title}{Displacement-driven approach to nonlocal elasticity}.
\newblock \emph{\bibinfo{journal}{Eur. J. Mech. A Solids}}
  \textbf{\bibinfo{volume}{92}}, \bibinfo{pages}{104434}
  (\bibinfo{year}{2022}).

\bibitem{schuster1997bidirectional}
\bibinfo{author}{Schuster, M.} \& \bibinfo{author}{Paliwal, K.~K.}
\newblock \bibinfo{title}{Bidirectional recurrent neural networks}.
\newblock \emph{\bibinfo{journal}{IEEE Trans. Signal Process.}}
  \textbf{\bibinfo{volume}{45}}, \bibinfo{pages}{2673--2681}
  (\bibinfo{year}{1997}).

\bibitem{sidhardh2020geometrically}
\bibinfo{author}{Sidhardh, S.}, \bibinfo{author}{Patnaik, S.} \&
  \bibinfo{author}{Semperlotti, F.}
\newblock \bibinfo{title}{Geometrically nonlinear response of a
  fractional-order nonlocal model of elasticity}.
\newblock \emph{\bibinfo{journal}{Int. J. Non Linear Mech.}}
  \textbf{\bibinfo{volume}{125}}, \bibinfo{pages}{103529}
  (\bibinfo{year}{2020}).

\bibitem{challamel2008small}
\bibinfo{author}{Challamel, N.} \& \bibinfo{author}{Wang, C.}
\newblock \bibinfo{title}{The small length scale effect for a non-local
  cantilever beam: a paradox solved}.
\newblock \emph{\bibinfo{journal}{Nanotechnology}}
  \textbf{\bibinfo{volume}{19}}, \bibinfo{pages}{345703}
  (\bibinfo{year}{2008}).

\bibitem{patnaik2020towards}
\bibinfo{author}{Patnaik, S.}, \bibinfo{author}{Sidhardh, S.} \&
  \bibinfo{author}{Semperlotti, F.}
\newblock \bibinfo{title}{Towards a unified approach to nonlocal elasticity via
  fractional-order mechanics}.
\newblock \emph{\bibinfo{journal}{Int. J. Mech. Sci.}}
  \textbf{\bibinfo{volume}{189}}, \bibinfo{pages}{105992}
  (\bibinfo{year}{2021}).

\bibitem{ding2021multiscale}
\bibinfo{author}{Ding, W.}, \bibinfo{author}{Patnaik, S.} \&
  \bibinfo{author}{Semperlotti, F.}
\newblock \bibinfo{title}{Multiscale nonlocal elasticity: A distributed order
  fractional formulation}.
\newblock \emph{\bibinfo{journal}{Int. J. Mech. Sci.}}
  \textbf{\bibinfo{volume}{226}}, \bibinfo{pages}{107381}
  (\bibinfo{year}{2022}).
\end{thebibliography}
\end{document}